# An Open-Source Project for MapReduce Performance Self-Tuning

Donghua Chen

*Abstract*—Many Hadoop configuration parameters have significant influence in the performance of running MapReduce jobs on Hadoop. It is time-consuming and tedious for general users to manually tune the parameters for optimal MapReduce performance. Besides, most of existing self-tuning system have opaque implementation, making it difficult to use in practice. This study presents an open-source project that hosts the developing self-tuning system called Catla to address the issues. Catla integrates multiple direct search and derivative-free optimization-based techniques to facilitate tuning efficiency for users. An overview of the system and its usage are illustrated in this study. We also reported a simple example demonstrating the benefits of this ongoing project. Although this project is still developing and far from comprehensive, it is dedicated to contributing Hadoop ecosystem in terms of improving performance in big data analysis.

## I. Introduction

THE performance of MapReduce jobs on Hadoop relies on settings of proper Hadoop configuration parameters [1]. Improper use of these parameters results in poor MapReduce performance, affecting users' actual business applications [2, 3]. The manual tuning process for improving performance of MapReduce jobs by experts is complicated and time-consuming even for experts, not to mention general users [4]. Therefore, it is necessary to improve such processes in a self-tuning manner to rapidly obtain the optimal values of multiple Hadoop parameters for MapReduce jobs.

Few of self-tuning systems [5-7] for Hadoop are open-source, making it difficult to utilize in real life and evaluate in scientific research. Starfish [8] has been outdated due to rapid iteration of Hadoop versions at present. Some systems used machine learning-based techniques to facilitate their tuning performance, which are state-of-art in this field [9, 10], which is difficult to evaluate due to their opaque implementation. Besides, it is difficult for general users to execute a MapReduce job and obtain metrics of performance after job completion. Besides, few systems can tune the performance using black-box optimization techniques to address the issue of noisy of running time of MapReduce jobs due to dynamic and complicated context of Hadoop cluster.

This study presents a basic overview of the open-source project called Catla [11] to address the issues in a more flexible and simpler manner. Design of self-tuning approaches in Catla

Manuscript submitted on December 28, 2019.
D. Chen is with the Department of Information Management, School of Economics and Management, Beijing Jiaotong University, Beijing, China.
*Corresponding Author: Donghua Chen, dhchen@bjtu.edu.cn

is not illustrated in the present study. Overall, Catla uses rule-based templates to organize necessary information of tuning MapReduce jobs and organize their tuning historical logs in a consistent way that can utilize in visualizing, predicting, optimizing and improving performance of MapReduce jobs on Hadoop.

## II. Methods

In this section, the architecture of Catla to run on Hadoop is firstly illustrated. Then, the setup environment of Catla is illustrated. Finally, the algorithms used in the proposed self-tuning system to improve tuning efficiency are briefly introduced.

### A. Architecture

Figure 1 describes the architecture of Catla to facilitate tuning efficiency of Hadoop configuration parameters in a flexible and automated manner, which solves the problem in the traditional tuning process.

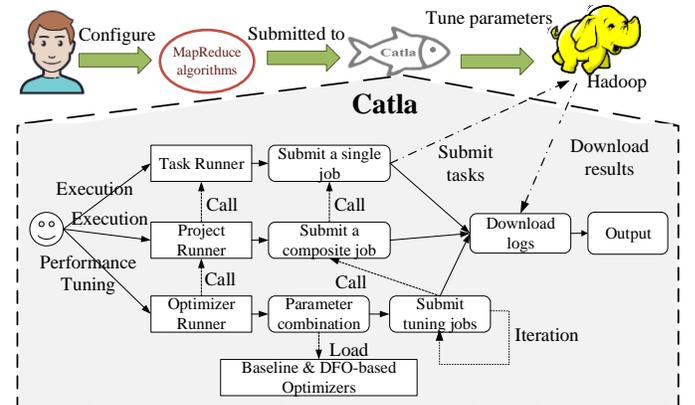

Fig. 1. Architecture of Catla

Catla has three components, namely, *Task Runner*, *Project Runner* and *Optimizer Runner*.

The *Task Runner* submits a single MapReduce job to a Hadoop cluster and obtains its analyzing results and logs after the job is completed. This component provides the basis of *Project Runner* and *Optimizer Runner*.

The *Project Runner* submits a group of MapReduce jobs in an organized project folder and monitor the status of its running until job completion; eventually, all analyzing results and their logs that contain information of running time in all MapReduce phrases are downloaded and organized to specified location in its project folder.

The *Optimizer Runner* creates a series of MapReduce jobs with different combinations of parameter values according to parameter configuration files and obtains the optimal parameter value sets with minimum running time after the tuning process is finished.

Two types of self-tuning processes using derivative-free optimization (DFO) techniques [12] and direct search techniques [13] respectively are planned to supported in Catla.

*B. Tuning Environment*

*1) Premise*

Tuning MapReduce job performance on Catla successfully must satisfy the premise of some environment settings. First, Catla must run on a computer located in the same network as a Hadoop cluster. Second, all job operations are performed through Secure Shell (SSH) clients to access master host in the cluster. Then, the current version of Catla is built on Hadoop 2.7.2, which may be possible to run on any Hadoop 2.x.x cluster. The computer that runs Catla must have properly installed Java environment. Finally, the log aggregation of Yarn must enable for proper retrieval of historic job logs after job completion.

*2) Tuning Process*

Catla employs a simple way to run tuning processes for MapReduce jobs. Here is a simple running process to run a task-based template project for submitting a MapReduce jobs and then obtaining analyzing results of MapReduce jobs as follows.

Step 1: Prepare for the executable binary *Catla.jar* stored in your workstation in the same network as the target Hadoop cluster, and the folder of a tuning project based on the project- and tuning-based templates.

Step 2: Change the master host's information defined in 'HadoopEnv.txt' from the project folder according to the users' actual Hadoop cluster.

Step 3: Open a Windows Command program, change current directory into the task-based folder downloaded Catla's GitHub repository.

Step 4: Run the Java Command '*java -jar Catla.jar -tool task -dir task_wordcount*' where the 'task_wordcount' is the relative path of a project folder.

Step 5: After the job is finished, the 'task_wordcount' folder should create a new folder 'downloaded_results' which stores the analyzing result of the job.

The steps are a simple demonstration example; more advanced examples for different tuning purposes will be released as peer-reviewed papers in the future.

*C. Tuning Approach*

This section provides the descriptions of tuning approaches in Catla to execute, monitor and tuning MapReduce in a Hadoop cluster. Key tuning approaches int the system are as follows:

*1) MapReduce job execution*

This task is to submit a MapReduce job within a Java library (jar file) to Hadoop cluster and obtain Hadoop log files and download results to a local computer after job completion.

*2) Direct search methods*

This task is to tune the performance of MapReduce jobs using exhaustive search, which means the system tries all combinations of parameter values to test the job and obtain a summary of running time vs. parameter values after the tuning process is finished.

*3) DFO-based search methods*

Tuning performance of MapReduce jobs using DFO methods, several DFO optimizers are being integrated in Catla.

*4) Log aggregation*

When the tuning process is stopped in the middle of tuning, the log aggregation is not finished. Therefore, the user can start this command to re-aggregate existing logs from /history folder.

*5) Visualization*

After job completion, the summaries of job metrics are in the sub folder "/history" of the project root folder that you run. Then, you can visualize the results from the information of *.csv files in the history folder by using statistics software such as Minitab and MATLAB software.

*D. CatlaUI*

To simply the complexity of the tuning process, a simple desktop version (CatlaUI) built upon core libarries of Catla is developed. The desktop version provides user-friendly interfaces for general users to run, monitor and tune a MapReduce without Windows commands.

Its features include: (1) it supports all commands of Catla in a user-friendly interface using SWT; (2) For DFO-based optimizers, CatlaUI provides a line chart to demonstrate change of running time of MapReduce jobs over number of iterations.

## III. RESULTS

A simple experiment using Catla is conducted here. The experiment aims to obtain optimal parameter values of Hadoop configuration with minimum running time of the MapReduce job, WordCount.

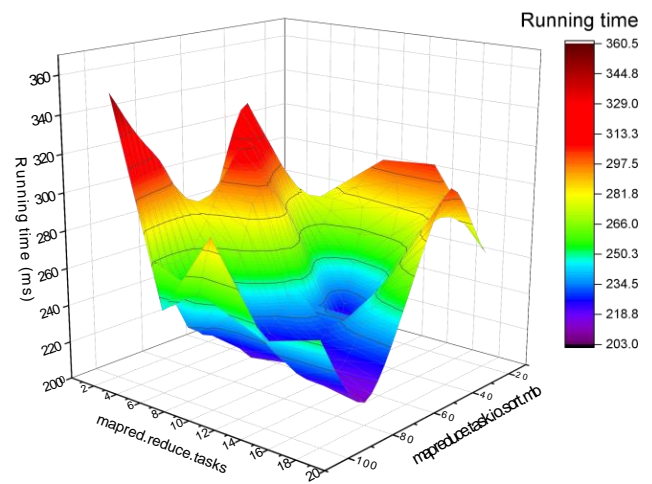

Fig. 2. Change of running time of the WordCount MapReduce job over two parameters, namely, *map.reduce.tasks* and *mapreduce.task.io.sort.mb*.

Here we focused on two parameters, namely, *map.reduce.tasks* and *mapreduce.task.io.sort.mb*. Figure 2 showed the change of running time of the MapReduce job over the parameters. The figure illustrates huge fluctuations in performance over the parameters *map.reduce.tasks* and *mapreduce.task.io.sort.mb* where larger *reduce.task* and larger *io.sort.mb* tends to reduce running time of the job. This simple experiment demonstrates the complexity of tuning processes for obtaining best performance of MapReduce jobs. The exhaustive search method for parameters can easily create a three-dimensional surface plots like Fig. 2 while the Catla system provides advanced optimization techniques to reduce tuning time like the use of DFO-optimizers.

Figure 3 illustrates the change of running time of a MapReduce job over number of iterations. The figure showed that with the use of the BOBYQA method, which is a DFO-based optimizer, has trends of convergence, the method can quickly obtain a stable minimum value of running time. The parameter values for corresponding minimum running time are considered as an optimal solution for the MapReduce job.

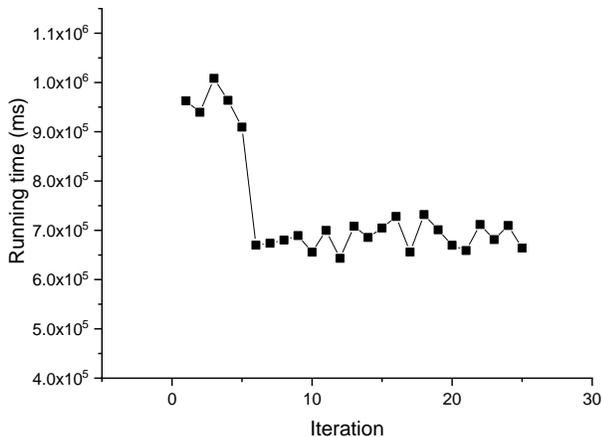

Fig. 3. Comparison of change of running time over number of iterations using the BOBYQA optimizer

## IV. RELATED WORK

Many existing self-tuning frameworks have been proposed to address specific problems in Hadoop parameter tuning. MRTune [14] is the first to consider data skew and task failures. It can work well when data are skewed. Cheng et al. [15] proposed a self-adaptive task tuning approach, namely, Ant, which automatically searches for the optimal configurations of individual tasks running on different nodes. Bei et al. [16] introduced MapreducE Self-Tuning (MEST) to accelerate the search process of Hadoop parameter configuration by integrating a model tree algorithm with a genetic algorithm. MEST significantly reduces search time by eliminating unnecessary profiling, modeling, and searching steps. A history-based auto-tuning (HAT) MapReduce scheduler is introduced to tune the weight of each phase of a map task and a reduce task in accordance with their values in history tasks. HAT uses the accurate weights of the phases to calculate the progress of current tasks. AutoTune [17] constructs a small-scale test bed for a production system to generate an additional sample and train a better model for optimizing application execution time on a big data analytics framework.

However, few of these systems provides public implementation or software that can be easily used in practice. Our study aims to solve this issue. More methodological papers inside the proposed Catla system are being peer-reviewed at the time being and will be released when they are accepted.

## V. CONCLUSIONS

This study proposes the basic overview of an open-source self-tuning system called Catla for MapReduce performance tuning based on multiple search methods. The proposed system provides a non-invasive and flexible scheme for tuning MapReduce performance without redesigning existing Hadoop clusters. The direct search methods and the DFO-based search methods are utilized to improve the tuning efficiency of parameters for the optimal performance of MapReduce jobs. The Catla project is still developing and will publish part of code implementation of tuning approaches in Catla via GitHub for scientific research.